\documentclass[a4paper,pra,twocolumn,superscriptaddress,showpacs]{revtex4}

\pdfoutput=1
\usepackage{amssymb}
\usepackage{amsmath}
\usepackage{epsfig}
\usepackage{color}
\usepackage{graphics, graphicx}
\usepackage{bbold}
\usepackage{psfrag}
\usepackage{mathcomp}
\usepackage{subfigure}
\usepackage{verbatim}
\usepackage{color}
\usepackage[colorlinks,citecolor=blue]{hyperref}

\begin{document}

\date{\today}
\title{Effective Hamiltonians for quasi-one-dimensional Fermi gases with spin-orbit coupling}
\author{Ren Zhang}
\author{Wei Zhang}
\email{wzhangl@ruc.edu.cn}

\affiliation{Department of Physics, Renmin University of China,
Beijing 100872, People's Republic of China.}

\begin{abstract}
We derive one-dimensional effective Hamiltonians for spin-orbit coupled Fermi gases confined in quasi-one-dimensional
trapping potentials. For energy regime around the two-body bound state energy, the effective Hamiltonian takes a two-channel
form, where the population in transverse excited levels are described by dressed molecules in the closed channel.
For energy regime slightly above the continuum threshold, the effective Hamiltonian takes a single-channel form, where
low-energy physics is governed by the one-dimensional interacting strength determined by three-dimensional scattering
length and transverse confinement. We further discuss the effect of spin-orbit coupling and effective Zeeman field on
the position of confinement-induced resonances, and show that these resonances can be understood as Feshbach
resonances between the threshold of the transverse ground state and the two-body bound state associated with 
the transverse excited states.
We expect that the shift of confinement-induced resonances can be observed under present experimental technology
at attainable temperatures.
\end{abstract}

\maketitle

\section{introduction}
The recent experimental realizations of synthetic spin-orbit coupling (SOC) in ultracold Fermi gases have paved a new route
towards the investigation of exotic phenomena in spin-orbit coupled Fermi systems~\cite{lin-11, wang-12,cheuk-12}.
A considerable amount of effort has been devoted to the characterization of multiple exotic superfluid phases, which
can emerge in various configurations of dimensionality and forms of SOC~\cite{iskin-11, gong-11, yu-11, hu-11,
zhang-08,sato-09,zhou-11, yi-12, yang-12, gong-12, he-12, han-12, wu-13, iskin-13, zhou-13, dong-13, qu-13, zhang-13, liu-12}.
Among these works, the investigation in low-dimensional systems is of particular interest,
partly due to the possibility to stabilize various exotic, in many cases topologically nontrivial, pairing states.
Specifically, in a two-dimensional (2D) Fermi gas with Rashba SOC
and out-of-plane Zeeman field, it has been shown that a topological superfluid state can be stabilized, and support topologically
protected edge states and Majorana zero modes at the cores of vortex 
excitations~\cite{zhang-08,sato-09,zhou-11, yi-12, yang-12, gong-12, he-12}.
For a 2D Fermi gas with a one-dimensional SOC which has been realized at National Institute of Standard and Technology 
(NIST)~\cite{lin-11}, nodal superfluid states with gapless excitations exist \cite{han-12, wu-13, iskin-13}.
Furthermore, it has been suggested that for a 2D Fermi gas under NIST SOC and both in-plane and out-of-plane Zeeman fields,
finite center-of-mass momentum Fulde-Ferrell-Larkin-Ovchinnikov (FFLO) states become the ground state of the system \cite{wu-13}.
These FFLO states are different from the conventional FFLO pairing states in a polarized Fermi gas in that they are driven by
SOC-induced spin mixing and transverse-field-induced Fermi surface deformation~\cite{zhou-13, dong-13, qu-13, zhang-13}.
In one dimensional (1D) Fermi gases, it has been also proposed that a topological superfluid
phase can be stabilized under NIST-type SOC, where Majorana modes are present at the edge of the system~\cite{liu-12}.

In the context of cold atomic gases, the physics in low dimensions is usually studied in quasi-low-dimensional configurations,
where strong confinements are applied along one or two spatial dimensions and the low-energy physics of the underlying system
can be safely described by an effective low-dimensional Hamiltonian. However, such a low-dimensional effective Hamiltonian
can not be obtained by simply projecting out all excited states along the strongly confined directions. Indeed, since the interatomic
interaction potential range is usually orders of magnitude smaller than the characteristic length scale of strong confinement,
the scattering process between two particles inevitably involves these excited states and is three-dimensional (3D) in nature.
As a consequence, one has to take into account all degrees of freedom and renormalize them in a proper way to extract a
correct effective Hamiltonian.

One way to obtain such an effective Hamiltonian is by matching the two-body physics. The physical consideration behind
this procedure is that cold atomic systems are usually prepared in the dilute limit, where the properties of the system are
governed by two-body processes. By matching solutions of a given two-body state, one can obtain the effective
low-dimensional Hamiltonian which can mimic the original quasi-low-dimensional Hamiltonian within a certain range of energy
around the eigenenergy of the corresponding two-body state. Following this procedure, Olshanii~\cite{olshanii-98}
focus on the scattering states in quasi-1D geometry, and demonstrate that the scattering processes can be well
described by a 1D single-channel model, where the effective 1D scattering resonance can be tuned by either
changing the 3D scattering length or varying the transverse confinement,
where the latter scheme is then referred as confinement-induced  resonance (CIR). In a later stage, Kestner and
Duan~\cite{kestner} investigate the two-body bound states, and show that when the two-body binding energy
is comparable or exceeds the transverse confinement, a single-channel model is not sufficient to mimic the original system,
and one has to rely on a two-channel model with dressed molecular degrees of freedom. In the presence of SOC,
theoretical analysis has been achieved for quasi-2D configurations with Rashba SOC~\cite{peng, ren},
but is still lacking for quasi-1D cases with NIST-type SOC.

In this manuscript, we investigate two-body problems in a quasi-1D Fermi system
with in-line SOC and transverse Zeeman field, and obtain the effective 1D Hamiltonians for different energy regimes.
For energy below threshold, we match solutions of two-body bound states and obtain the effective Hamiltonian
in a two-channel model form. This 1D model is suitable for the investigation on pairing phases in quasi-1D Fermi
gases with SOC, for the fermionic chemical potential in such a case is close to one half of the two-body bound state
energy in the dilute limit. For energy around threshold, we match the zero-energy scattering amplitude and obtain the
effective Hamiltonian in a single-channel form. We further analyze the effect of SOC and Zeeman field on
the position of confinement-induced resonances, which can be used to create 1D systems in strongly-interacting limit.
We note that with the present technology, such a system is experimentally achievable by implementing
anisotropic traps or 2D optical lattices together with NIST-type SOC.

The reminder of this manuscript is organized as follows. In Sec.~\ref{sec:model}, we present our
formalism for two-body problems in quasi-1D configuration with NIST-type SOC. 
In Sec.~\ref{sec:bound states}, we analyze the two-body bound states.
Using these results, we then introduce an effective 1D Hamiltonian which is
suitable to investigate low-energy physics around two-body bound state energy. For scattering states
at threshold, we also derive an effective 1D Hamiltonian in Sec.~\ref{sec:scattering states}, and determine the position
of CIR in the presence of SOC and effective Zeeman fields in Sec.~\ref{sec:cir}.
Finally, we summarize in Sec.~\ref{sec:con}.

\section{two-body problems with NIST-type SOC in quasi-1D trapping potential}
\label{sec:model}

We consider two spin-$1/2$ fermionic atoms with one-dimensional SOC in a quasi-1D configuration.
The free Hamiltonian of this two-body system is given by
\begin{eqnarray}
\label{eqn:H}
{\hat H}_0 &=& -\frac{\hbar^2}{2m}\left(\nabla^{2}_{{\bf r}_{1}}+\nabla^{2}_{{\bf r}_{2}}\right)
+U({\bf r}_{1})+U({\bf r}_{2})
\nonumber\\
&& + {\hat H}_{\rm SOC}^1+ {\hat H}_{\rm SOC}^2,
\end{eqnarray}
where $m$ is the atomic mass, ${\bf r}_{j=1,2} \equiv (x_j, y_j, z_j)$ denotes the spatial position
of the $j$-th atom, and $U({\bf r}_j) \equiv m \omega^2 (y_j^2 + z_j^2)/2$ is the quasi-1D
confinement within the radial $y$-$z$ plane with trapping frequency $\omega$.
The SOC term takes the form
\begin{eqnarray}
\label{eqn:Hsoc}
{\hat H}_{\rm SOC}^j  = \lambda p_x^{(j)} \sigma_x^{(j)}  + h \sigma_z^{(j)}  + h_x \sigma_x^{(j)} ,
\end{eqnarray}
where $p_x^{(j)}$ is the linear momentum along the $x$-axis of the $j$-th atom,
$\sigma_i^{(j)}$ are Pauli matrices, $\lambda$ is the intensity of in-line SOC,
$h$ and $h_x$ are the effective Zeeman fields along the transverse and axial directions,
respectively. We notice that such type of SOC has been successfully implemented in Fermi systems
using the NIST scheme, where the transverse Zeeman field $h$ and in-line field $h_x$ correspond to
the Rabi frequency of Raman lasers $\Omega$ and the two-photon detuning $\delta$,
respectively~\cite{wang-12, cheuk-12}. Since the in-plane field $h_x$ is not crucial
for generating the topological superfluid phases~\cite{liu-12}, in our manuscript we consider
only the case of $h_x = 0$ for simplicity. However, we stress that all derivation
and calculation can be easily generated to cases with finite $h_x$.

Due to the presence of SOC, the center-of-mass (CoM) and relative degrees of freedom
along the SOC direction of the two-particle system is not separable. In fact, by representing
the quantum state of relative degree of freedom by a spinor wave function
\begin{eqnarray}
\label{eqn:wavefunction}
|\psi({\bf r})\rangle &=&
\psi_{\uparrow\uparrow}({\bf r})|\uparrow\rangle_1|\uparrow\rangle_2
+\psi_{\uparrow\downarrow}({\bf r})|\uparrow\rangle_1|\downarrow\rangle_2 \nonumber\\
& &
+\psi_{\downarrow\uparrow}({\bf r})|\downarrow\rangle_1|\uparrow\rangle_2
+\psi_{\downarrow\downarrow}({\bf r})|\downarrow\rangle_1|\downarrow\rangle_2,
\end{eqnarray}
the relative motion along the $x$-axis depends on the corresponding CoM momentum $Q_x$,
leading to the resulting Hamiltonian
\begin{eqnarray}
\label{eqn:rel-hamilton}
\hat{H}_{\rm rel} &=& -\sum_{s=x,y,z}\frac{\partial^2}{\partial s^2}
+\frac{1}{4}\sum_{s=y,z}(s^2-1) \nonumber\\
&& + \lambda\sum_{j=1,2}\left[ \frac{Q_x}{2}+(-1)^{j} k_x \right]\sigma_{x}^{(j)} + h\sum_{j=1,2}\sigma_{z}^{(j)}.
\end{eqnarray}
where $k_x$ is the relative momentum along the $x$-axis. 
Here, we adopt the natural unit with $\hbar=m=1$, and
set $\omega = 1$ as the energy unit. We also drop the zero-point energy of the transverse motion
along the $y$ and $z$ directions for simplicity, since the relative and CoM degrees of freedom
remain separable along these axes. Without loss of generality, we assume that the SOC intensity
$\lambda$ and the effective Zeeman field $h$ are real and positive definite.

The interaction effect between the two particles can be analyzed by implementing the
Bethe-Peierls boundary condition in three dimensions
\begin{eqnarray}
\label{eqn:bp1}
\psi\propto(1/r-1/a_s),\quad r\rightarrow0.
\end{eqnarray}
The validity of this boundary condition can be understood by noticing the separation of energy
or length scales. In fact, The length scale associated with the inter-particle interacting
potential $R_e$ is in the scale of $10^{-9}$m. As a comparison, the characteristic length scales
associated with the transverse trap $a_\perp$ and the spin-orbit coupling
$a_\lambda \equiv 1/\lambda $ are both in the scale of $10^{-6}$m, which are several orders
of magnitude larger than $R_e$. Thus, the presence of a transverse trap and SOC will not alter
the divergence behavior of scattering wave functions. For sufficiently low energy, i.e. $kR_e\ll1$,
where $k$ is the relative wave vector between two particles, the scattering is dominated by $s$-wave contribution,
and the behavior of the wave function at the inter-particle  distance $r\gg R_e$ is determined by the
3D $s$-wave scattering length $a_s$, which is related to the scattering phase
shift as $\delta_s = - \arctan(k a_s)$.
Hence, to obtain the correct expression for the wave function at distance $r\gg R_e$, which is the regime
we are interested in, we need to solve the Shr$\ddot{\rm o}$dinger equation for the relative motion
Eq. (\ref{eqn:rel-hamilton}) under the boundary condition Eq. (\ref{eqn:bp1}).

To investigate the two-body scattering process in the presence of SOC, we first define the single-particle
spin state for the $j$-th atom
\begin{eqnarray}
\label{eqn:single-spin-state}
(t_x\sigma_x^{(j)}+t_z\sigma_z^{(j)})|\alpha_j,\textbf{t}\rangle \equiv \alpha_j |t| |\alpha,\textbf{t}\rangle,
\end{eqnarray}
where ${\bf t}=(t_x,t_z)$ is a two-dimensional vector in the $x$-$z$ plane, and $\alpha_j=\pm1$ denotes
the helicity index. From now on, we focus on the zero CoM momentum case with $Q_x=0$ as a particular
example, and note that the same procedure can be easily extended to cases with finite $Q_x$.
The two-particle spin state can be defined as
\begin{eqnarray}
|{\bm \alpha}(k_x)\rangle=|\alpha_1,\left(\lambda k_x,h\right)\rangle_1|\alpha_2,\left(-\lambda k_x,h\right)\rangle_2
\end{eqnarray}
with ${\bm \alpha} \equiv (\alpha_1,\alpha_2)$ the shorthand notation.
The incident wave function with the proper symmetry thus can be expressed in the following form
\begin{eqnarray}
\label{eqn:incident-function}
|\psi^{(0)}_c(\textbf{r})\rangle=\frac{e^{ik_xx}}{2\sqrt{\pi}}\left[|{\bm \alpha}(k_x)\rangle
-|{\bm \alpha}'(-k_x)\rangle\right]
\phi_0(y)\phi_0(z),
\end{eqnarray}
where $\phi_0$ is the ground state of a one-dimensional harmonic oscillator,
and ${\bm \alpha}' \equiv (\alpha_2,\alpha_1)$. In the rest of this paper, we denote
the scattering channel by $c=({\bm \alpha},k_x)$. The eigen-energy corresponding to
the incident wave function can be obtained by a straightforward calculation, leading to
\begin{eqnarray}
\label{eqn:eigenvalue}
\varepsilon=\varepsilon_c+(m+n),
\end{eqnarray}
where $\varepsilon_c=k_x^2+\left(\alpha_1+\alpha_2\right)\sqrt{\lambda^2k_x^2+h^2}$.
One should notice that in the presence of SOC, the threshold energy is shifted from zero
to a nonzero value
\begin{eqnarray}
\label{eqn:threshold}
\varepsilon_{\rm th}=
\begin{cases}
-2h,\quad \lambda^2<h;\\
-\lambda^2-h^2/\lambda^2,\quad \lambda^2\geq h.
\end{cases}
\end{eqnarray}

As the scattering energy is low enough with $\varepsilon_c - \varepsilon_{\rm th} \ll 1/R_e^2$,
the wave function of the scattering state can be expressed as~\cite{peng, taylor, petrov01}
\begin{eqnarray}
\label{eqn:scattering state}
|\psi_c(\textbf{r})\rangle\approx|\psi_c^{(0)}(\textbf{r})\rangle+\frac{A(c)}{\phi_0(0)^2}G_{0}(\varepsilon_c;{\bf r,0})|0,0\rangle
\end{eqnarray}
in the asymptotic region with $r\gtrsim R_e$. Here, $A(c)$ is a coefficient to be determined,
 $G_0(\varepsilon_b;{\bf r},{\bf r}')$ is the Green's function associated with the free Hamiltonian which
describes the relative motion of the two colliding atoms
\begin{eqnarray}
\label{eqn:def-green}
G_0(\eta;{\bf r,r'})=\frac{1}{\eta+i0^+-\hat{H}_{\rm rel}}\delta({\bf r}-{\bf r'}),
\end{eqnarray}
and $|0,0\rangle$ represents a spin singlet state
\begin{eqnarray}
\label{eqn:singlet}
|0,0\rangle=\frac{1}{\sqrt{2}}\big(|\uparrow\rangle_1|\downarrow\rangle_2
- |\downarrow\rangle_1|\uparrow\rangle_2\big).
\end{eqnarray}

Similarly, when the energy $\varepsilon_b$ of the bound state is close enough to the
threshold with $\varepsilon_{\rm th}-\varepsilon_b\ll1/R_e^2$, the wave function
$|\psi_b({\bf r})\rangle$ of the two-body bound state can be approximated as
\begin{eqnarray}
\label{eqn:bound state}
|\psi_b(\textbf{r})\rangle\approx B G_0(\varepsilon_b;{\bf r,0})|0,0\rangle
\end{eqnarray}
in the region of $r\gtrsim R_e$,  where $B$ is the normalization constant.
The coefficients $A(c)$ and $B$ in Eqs.~(\ref{eqn:scattering state}) and (\ref{eqn:bound state})
can be derived by implementing the Bethe-Peierls boundary condition (\ref{eqn:bp1}).
Specifically, by using the identity
\begin{eqnarray}
\label{eqn:delta}
\delta({\bf r}-{\bf r'})&=&\int_{-\infty}^{\infty} dk_x\frac{e^{ik_x(x-x')}}{2\pi}\left(\sum_{\bm \alpha}|{\bm \alpha}(k_x)\rangle\langle{\bm \alpha}(k_x)|\right)\nonumber\\
&&\times
\left(\sum_{m,n}\phi^*_m(y')\phi_m(y)\phi^*_n(z')\phi_n(z)\right)
\end{eqnarray}
and the relation
\begin{eqnarray}
\label{eqn:eigenfunction}
&&\hat{H}_{\rm rel}e^{ik_xx}\phi_m(y)\phi_n(z)|{\bm \alpha(k_x)}\rangle
\nonumber \\
&& \hspace{1cm}
=(\varepsilon_c+m+n)e^{ik_xx}\phi_m(y)
\phi_n(z)|{\bm \alpha(k_x)}\rangle
\end{eqnarray}
with $\phi_m$ the $m^{\rm th}$ eigenstate of a one-dimensional harmonic oscillator,
the behavior of the Green's function $G_0(\eta;{\bf r,0})$ at $r \to 0$ can be obtained as
\begin{eqnarray}
\label{eqn:tot-green}
\langle0,0|G_0(\eta;{\bf r,0})|0,0\rangle&=&\langle0,0|g(\eta;{\bf r,0})|0,0\rangle + {\cal S}(\eta,{\bf r}).
\nonumber \\
\end{eqnarray}
The functions in the equation above are defined as
\begin{widetext}
\begin{eqnarray}
\label{eqn:green1}
\langle0,0|g(\eta;{\bf r,0})|0,0\rangle&=&\sum_{m,n}\int dk_x\frac{e^{ik_xx}}{2\pi}\frac{1}{\eta+i0^+-\left(k_x^2+m+n\right)}\phi^*_m(0)\phi_m(y)\phi^*_n(0)\phi_n(z),
\end{eqnarray}
\begin{eqnarray}
\label{eqn:s-function}
{\cal S}(\eta,{\bf r})&=&\sum_{{\bm \alpha'},m,n}\int_{-\infty}^{\infty} dk'_x\frac{e^{ik'_xx}}{2\pi}|\langle0,0|{\bm \alpha'}(k_x')\rangle|^2\phi^*_m(0)\phi_m(y)\phi^*_n(0)\phi_n(z)\nonumber\\
&&\times\left[\frac{1}{\eta+i0^+-(\varepsilon_{c'}+m+n)}-\frac{1}{\eta+i0^+-\left(k_x'^2+m+n\right)}\right].
\end{eqnarray}
\end{widetext}
The two terms on the right-hand-side of Eq. (\ref{eqn:tot-green}) can be understood as follows.
The first term is the Green's function associated with the relative Hamiltonian as SOC is absent, while
the second term is the contribution induced by the SOC. By substituting Eqs. (\ref{eqn:tot-green}-\ref{eqn:s-function})
into Eqs. (\ref{eqn:scattering state}) and (\ref{eqn:bound state}), the scattering and bound states
can be solved under the Bethe-Peierls boundary condition Eq. (\ref{eqn:bp1}), respectively.

\section{Bound states and the two-channel effective Hamiltonian}
\label{sec:bound states}

In this section, we focus on the energy regime below threshold and investigate the two-body
bound state with eigen-energy $\varepsilon_b$. From Eq. (\ref{eqn:green1}),
the behavior of $\langle0,0|g(\varepsilon_b;{\bf r,0})|0,0\rangle$
at vanishing $r$ reads
\begin{eqnarray}
\label{eqn:green-vinish-r}
\lim_{r\rightarrow0^+}\langle0,0|g(\varepsilon_b;{\bf r,0})|0,0\rangle
&=&-\frac{1}{2\pi \sqrt{-\varepsilon_b}}-{\cal F}(\varepsilon_b),
\end{eqnarray}
where ${\cal F}(\varepsilon_b)$ is defined as,
\begin{eqnarray}
\label{eqn:f-function}
{\cal F}(\varepsilon_b)&=&\frac{1}{2^{3/2}\pi}\lim_{x\rightarrow0^+}\sum_{s=1}^{\infty}\frac{e^{-\sqrt{s-\varepsilon_b/2}\sqrt{2}x}}{\sqrt{s-\varepsilon_b/2}}.
\end{eqnarray}
Here, we set the value of $\phi_0(0)$ in Eqs. (\ref{eqn:green1}) and (\ref{eqn:s-function})
to be real and positive, leading to the following expressions for
$|\phi_{2m}(0)|^2=\Gamma(m+1/2)/[\pi\Gamma(m+1)]$. Besides, we also utilize the identity
\begin{eqnarray}
\label{eqn:use-identity}
\sum_{n=0}^{s}\frac{\Gamma(s-n+1/2)\Gamma(n+1/2)}{\Gamma(s-n+1)\Gamma(n+1)}=\pi,\ {\rm for}\ s=1,2,3...
\end{eqnarray}
to obtain the final form of Eq. (\ref{eqn:f-function}).

The summation over $s$ in Eq. (\ref{eqn:f-function}) includes contribution from all transverse excited states,
and it cannot be interchanged with the limit of $x \to 0^+$ since the summation is not uniformly convergent.
By using the relation
\begin{eqnarray}
\label{eqn:identity}
\int_{0}^{\infty}
\frac{\exp(-\sqrt{s}\xi)}{\sqrt{s}} ds
=\sum_{s=1}^{\infty}\int_{s-1}^{s} \frac{\exp(-\sqrt{s'}\xi)}{\sqrt{s'}} d s'=\frac{2}{\xi},
\end{eqnarray}
we can finally obtain the behavior of $\langle0,0|g(\varepsilon_b;{\bf r,0})|0,0\rangle$ at vanishing $r$
\begin{eqnarray}
\label{eqn:green-vinish-r}
&&\lim_{r\rightarrow0^+}\langle0,0|g(\varepsilon_b;{\bf r,0})|0,0\rangle
\nonumber \\
&& \hspace{1cm}
= \frac{-1}{2\pi}\left[\frac{1}{r}-{\cal C}+\frac{\overline{\cal L}\left(-\varepsilon_b/2 \right)}{\sqrt{2}}+
\frac{1}{\sqrt{-\varepsilon_b}}\right],
\end{eqnarray}
where the function $\overline{\cal L}(\varepsilon)$ is defined as
\begin{eqnarray}
\label{eqn:l-function}
\overline{{\cal L}}(\varepsilon)&=&\sum_{n=1}^{\infty}(-1)^n\frac{\zeta(n+1/2)(2n-1)!!\varepsilon^n}{2^nn!}\nonumber\\
&=&\zeta\left[\frac{1}{2},1+\varepsilon\right]+\sqrt{2}{\cal C}
\end{eqnarray}
with ${\cal C} \approx 1.0326$ and $\zeta(s,a)$ is the Hurwitz Zeta function.

By using the identity Eq.~(\ref{eqn:use-identity}), we can evaluate the contribution to the Green's function
induced by the presence of SOC, leading to
\begin{eqnarray}
\label{eqn:s-function-1}
&&{\cal S}_b(\varepsilon_b) \equiv \lim_{r\rightarrow0^+}{\cal S} (\varepsilon_b,{\bf r})
= \frac{1}{2\pi}\sum_{s=0}^{\infty}\frac{b}{E_sb+d}
\nonumber \\
&& \times \left\{ \frac{\sqrt{2 \beta-2 E_s-b}
\left[2 d-b\left(\beta-E_s\right) \right]}{4 E_s b+b^2+4d}-\sqrt{-E_s}\right\},
\end{eqnarray}
where $\beta = \sqrt{E_s^2 -d}$, $E_s=\varepsilon_b - 2s$, $b=4\lambda^2$ and $d=4h^2$.
By substituting Eqs.~(\ref{eqn:green-vinish-r}) and (\ref{eqn:s-function-1}) into the
bound state wave function Eq. (\ref{eqn:bound state}) and the Bethe-Peierls boundary
condition Eq.~(\ref{eqn:bp1}), one can easily reach
\begin{eqnarray}
\label{eqn:Eb}
\frac{1}{a_s}&=&-\left(\frac{\zeta\left[1/2,1-\varepsilon_b/2 \right]}
{\sqrt{2}}+\frac{1}{\sqrt{-\varepsilon_b}}\right)+2\pi{\cal S}_b(\varepsilon_b).
\end{eqnarray}
Notice that the effect of SOC is represented in the last term on the right-hand-side of the equation above.
In fact, as the intensity of SOC goes to zero with $\lambda \to 0$, the function ${\cal S}_b$ vanishes
and we retrieve the equation for two-body bound states in the absence of SOC~\cite{olshanii-98, wzhang-11}.

With the knowledge of the two-body bound states we have discussed above, next we derive the
1D effective Hamiltonian, which can emulate the physics within the energy regime around the
two-body bound state energy $\varepsilon_b$. This effective Hamiltonian takes a two-channel model
form, where the open channel describes fermions residing on the transverse ground state, and the
closed channel is constructed by dressed molecules. The dressed molecules are considered as
structureless bosons, and are included to incorporate all high energy degrees of freedom including
Feshbach molecules and fermions on transverse excited states. In the presence of SOC and
effective Zeeman field, the effective Hamiltonian can be written as
\begin{eqnarray}
\label{eqn:two-interaction}
\hat{H}_{\rm eff}^{\rm 2c} = \hat{H}_{\rm eff}^{0} + \hat{H}_{\rm eff}^{\rm cc} + \hat{U}_{\rm eff},
\end{eqnarray}
where the free Hamiltonian and the atom-atom interaction are represented as,
\begin{eqnarray}
\hat{H}_{\rm eff}^{0} &=& \sum_{k,\sigma}\epsilon_k a_{k\sigma}^{\dagger}a_{k\sigma}
+ \lambda'\sum_{k}k (a_{k\uparrow}^{\dagger}a_{k\downarrow}+a_{k\downarrow}^{\dagger}a_{k\uparrow})\nonumber\\
&+&h'\sum_{k}(a_{k\uparrow}^{\dagger}a_{k\uparrow}-a_{k\downarrow}^{\dagger}a_{k\downarrow}).
\\
\hat{U}_{\rm eff} &=& \frac{V_b}{L} \sum_{k,k'}a_{k\uparrow}^{\dagger}
a_{-k\downarrow}^{\dagger}a_{-k'\downarrow}a_{k'\uparrow}.
\end{eqnarray}
Here, $a_{k\sigma}^{\dagger}$ ($a_{k\sigma}$) is the creation (annihilation) operator for
an atom with 1D momentum $k$ and spin $\sigma$, and $L$ is the quantization length of the system.
The threshold energy is chosen as $-1$ to match the zero-point energy of the two transverse
directions in the original quasi-1D configuration. The term involving the closed channel reads
\begin{eqnarray}
\hat{H}_{\rm eff}^{\rm cc} = \delta_bd^{\dagger}_0d_0+\frac{\alpha_b}{L^{1/2}}
\sum_{k}\left(a_{k\uparrow}^{\dagger}a_{-k\downarrow}^{\dagger}d_0+{\rm H.C.}\right),
\end{eqnarray}
where $d_0^{\dagger}$ and $d_0$ are the creation and annihilation operators for dressed molecules, respectively.
Here, $\delta_b$ is the detuning between open and closed channels, $\alpha_b$ denotes the atom-molecule interaction strength,
and H.C. stands for Hermitian conjugate.

The three bare parameters $V_b$, $\delta_b$ and $\alpha_b$ are related to the physical
ones via a 1D renormalization
\begin{eqnarray}
\label{eqn:renormalization}
&&V_c^{-1}=-\int \frac{d k}{(4\pi)}\frac{1}{\epsilon_k+1},\quad \Omega^{-1}=1+\frac{V_p}{V_c}\nonumber\\
&&V_p=\Omega^{-1} V_b,\quad \alpha_p=\Omega^{-1} \alpha_b,\quad \delta_p=\delta_b+\Omega\frac{\alpha_p^2}{V_c}.
\end{eqnarray}
Following the scheme in Ref.~\cite{ren}, we can reach
\begin{eqnarray}
\label{eqn:eff-main}
\left[V_b-\frac{\alpha_b^2}{\delta_b-\varepsilon_b}\right]^{-1}=2\pi\sigma_p(\varepsilon_b),
\end{eqnarray}
where the function $\sigma(\varepsilon_b)$ takes the form,
\begin{eqnarray}
\label{eqn:func-sigma}
\sigma(\varepsilon_b)&=&\frac{1}{2^{5/2}\pi}-\frac{\sqrt{\varepsilon_{b}^2-d}-\varepsilon_{b}}{4\pi\sqrt{2 \sqrt{\varepsilon_{b}^2-d}-2 \varepsilon_{b}-b}\sqrt{\varepsilon_{b}^2-d}}.\nonumber\\
\end{eqnarray}

The physical parameters in the effective Hamiltonian can be fixed by matching single- and two-body physics with
the original quasi-1D Hamiltonian. Specifically, we require the effective Hamiltonian to give the same single-particle
dispersion, two-body bound state energy, and the population of atoms in the transverse ground state as the original
model. Under these conditions, we get
\begin{eqnarray}
\label{eqn:parameter}
\lambda'&=&\lambda,
\nonumber \\
h'&=&h,\nonumber\\
V_p^{-1}&=&2\pi(U_p^{-1}-C_p),\nonumber\\
\delta_p&=&\varepsilon_b-\frac{\sigma_p(\varepsilon_b)}{\partial{\cal P}(\varepsilon_b)/\partial \varepsilon_b}
\left[1-\frac{\sigma_p(\varepsilon_b)}{U_p^{-1}-C_p}\right],\\
\alpha_p^2&=&\frac{1}{2\pi \partial{\cal P}(\varepsilon_b)/\partial \varepsilon_b}
\left[1-\frac{\sigma_p(\varepsilon_b)}{U_p^{-1}-C_p}\right]^2.\nonumber
\end{eqnarray}
Here, the parameters are defined as
\begin{eqnarray}
\label{eqn:parameter2}
C_p&=&{\cal S}_p(\varepsilon_b^{\rm inf})-\sigma_p(\varepsilon_b^{\rm inf}),\nonumber\\
{\cal P}(\varepsilon_b)&=&1/V_{\rm eff}-{\cal S}_p(\varepsilon_b)+\sigma_p(\varepsilon_p),\\
V_{\rm eff}&=&U_p-\frac{g_p^2}{\nu_p-E}\nonumber,
\end{eqnarray}
and $\varepsilon_b^{\rm inf}$ denotes the two-body bound state energy in quasi-one dimension for $\nu_p\rightarrow\infty$.
\begin{figure}
\includegraphics[width=4cm]{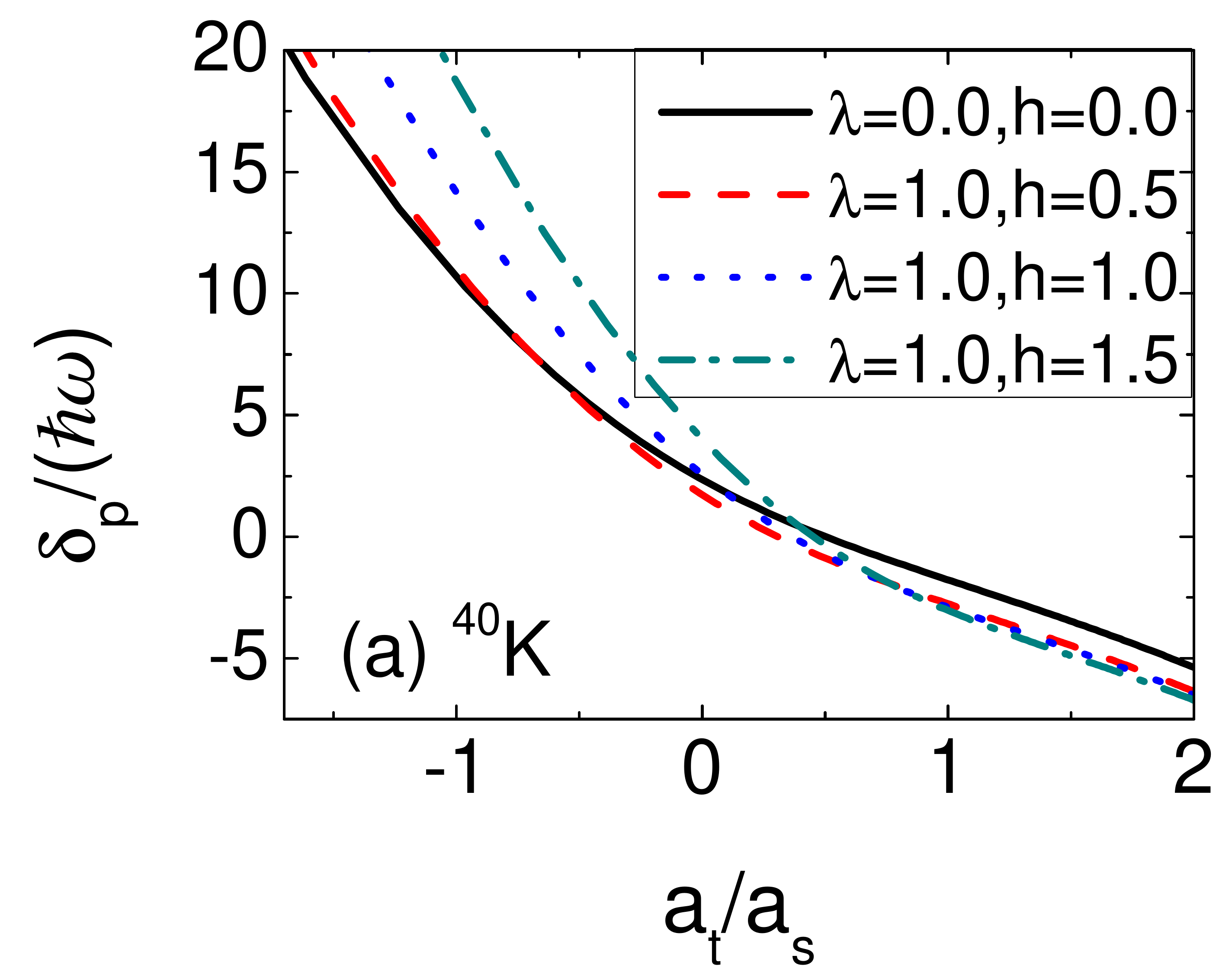}
\includegraphics[width=4cm]{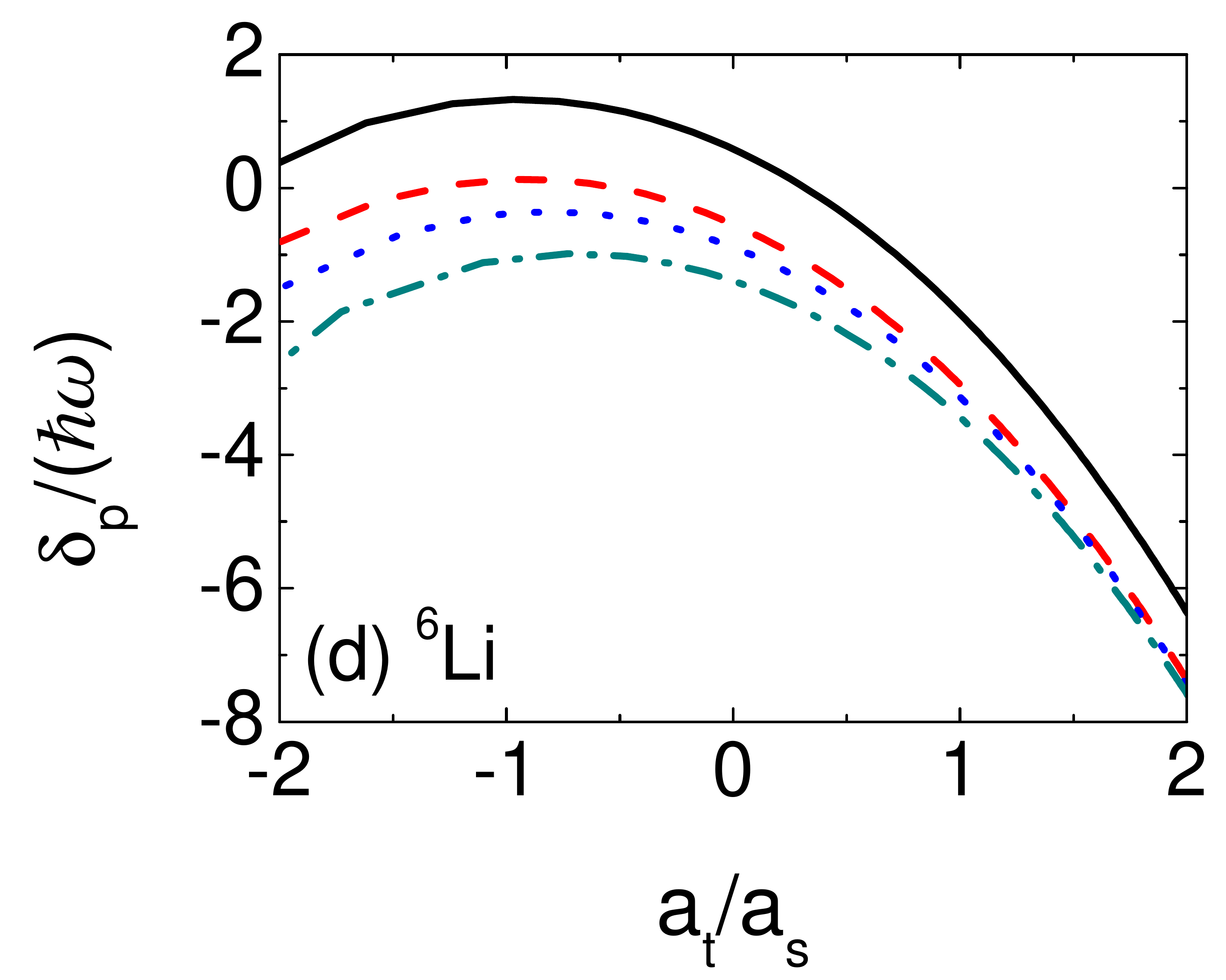}
\includegraphics[width=4cm]{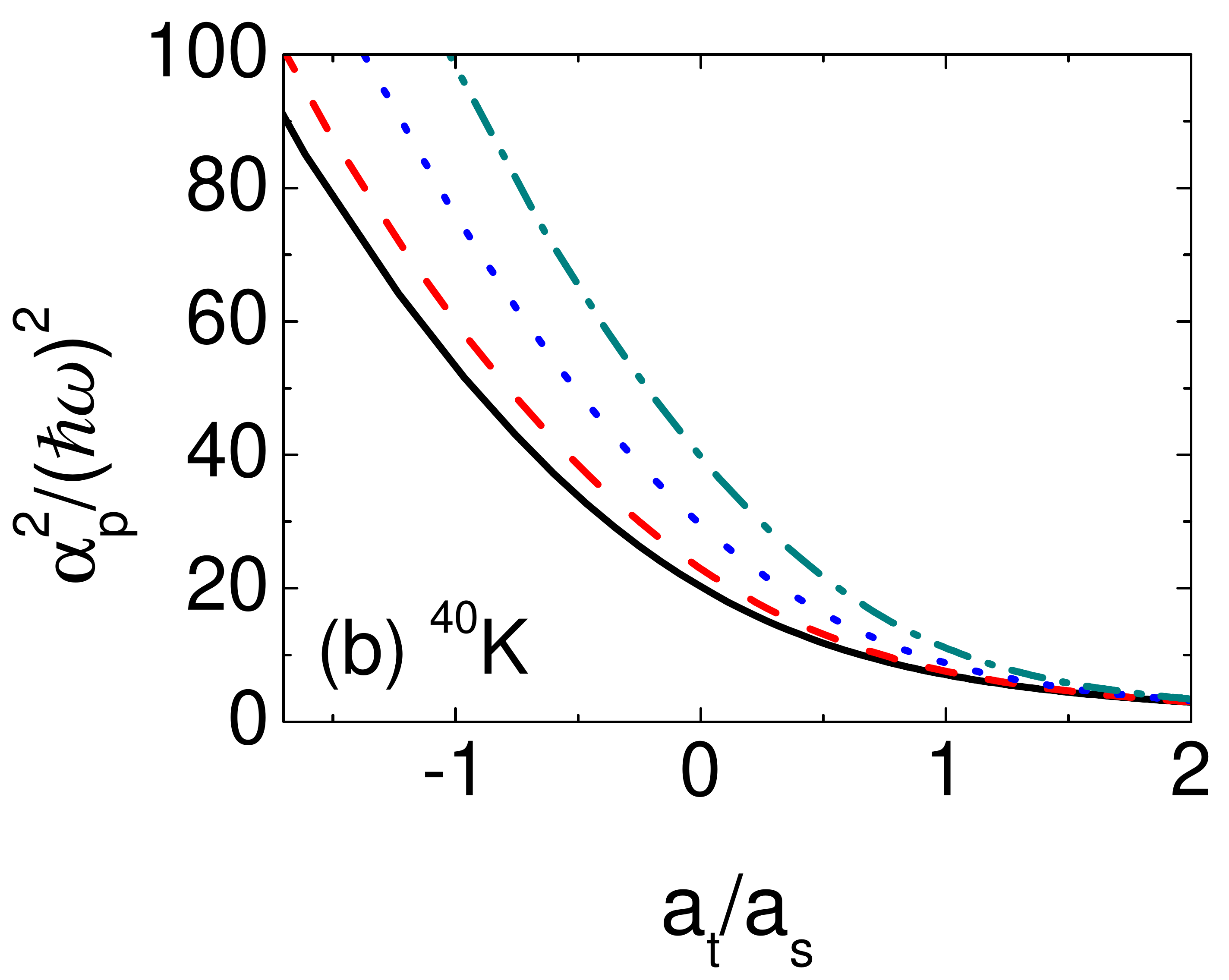}
\includegraphics[width=4cm]{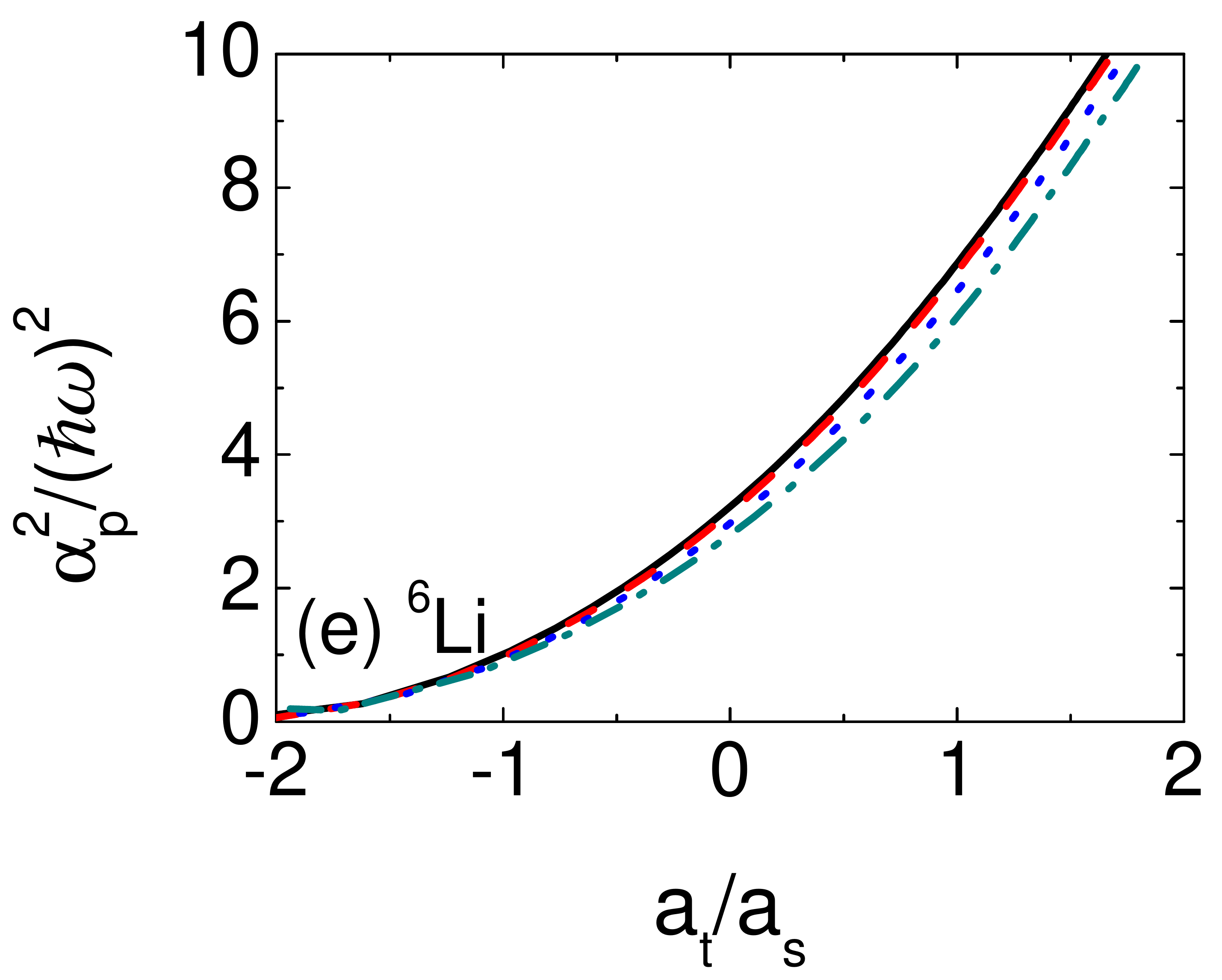}
\includegraphics[width=4cm]{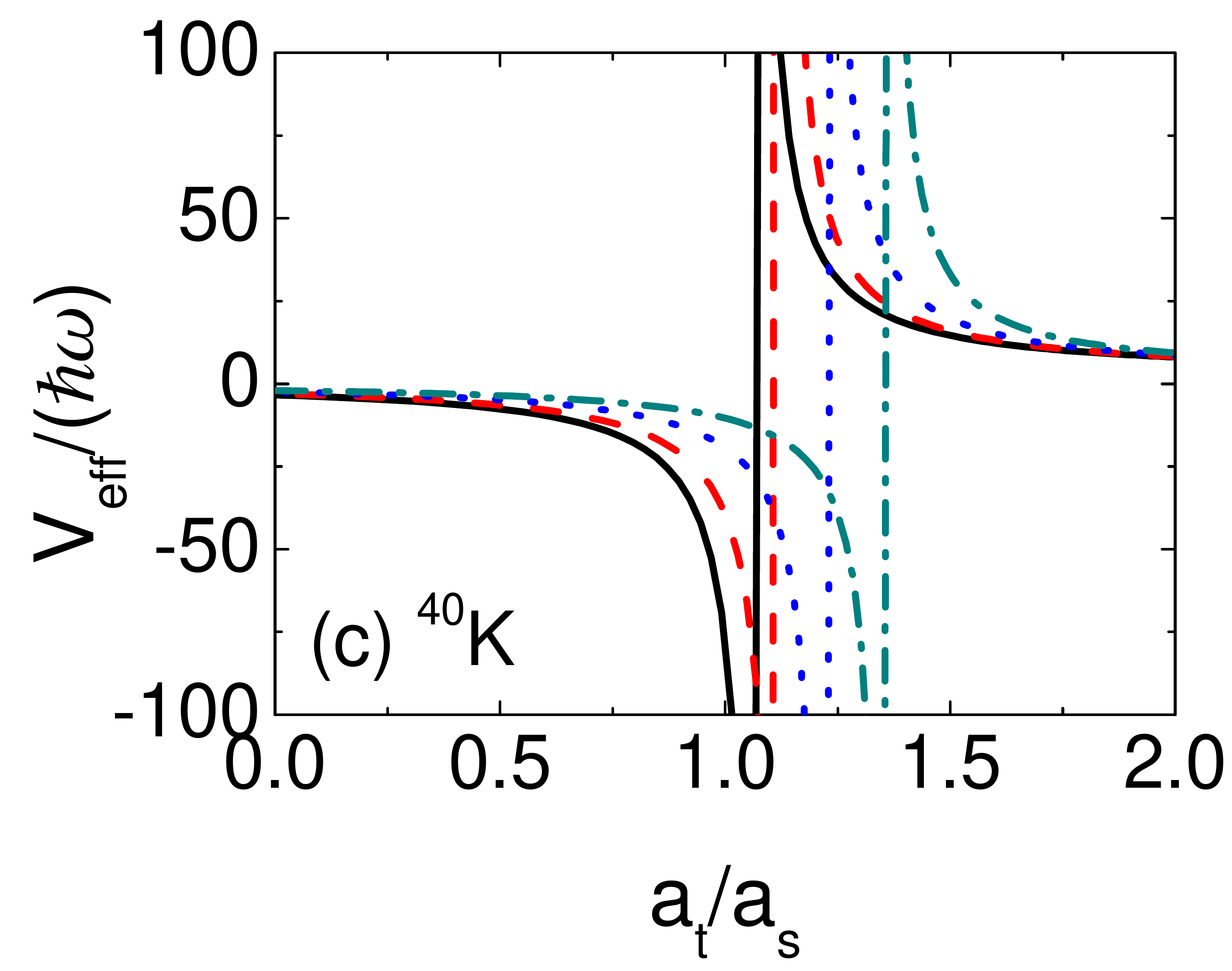}
\includegraphics[width=4cm]{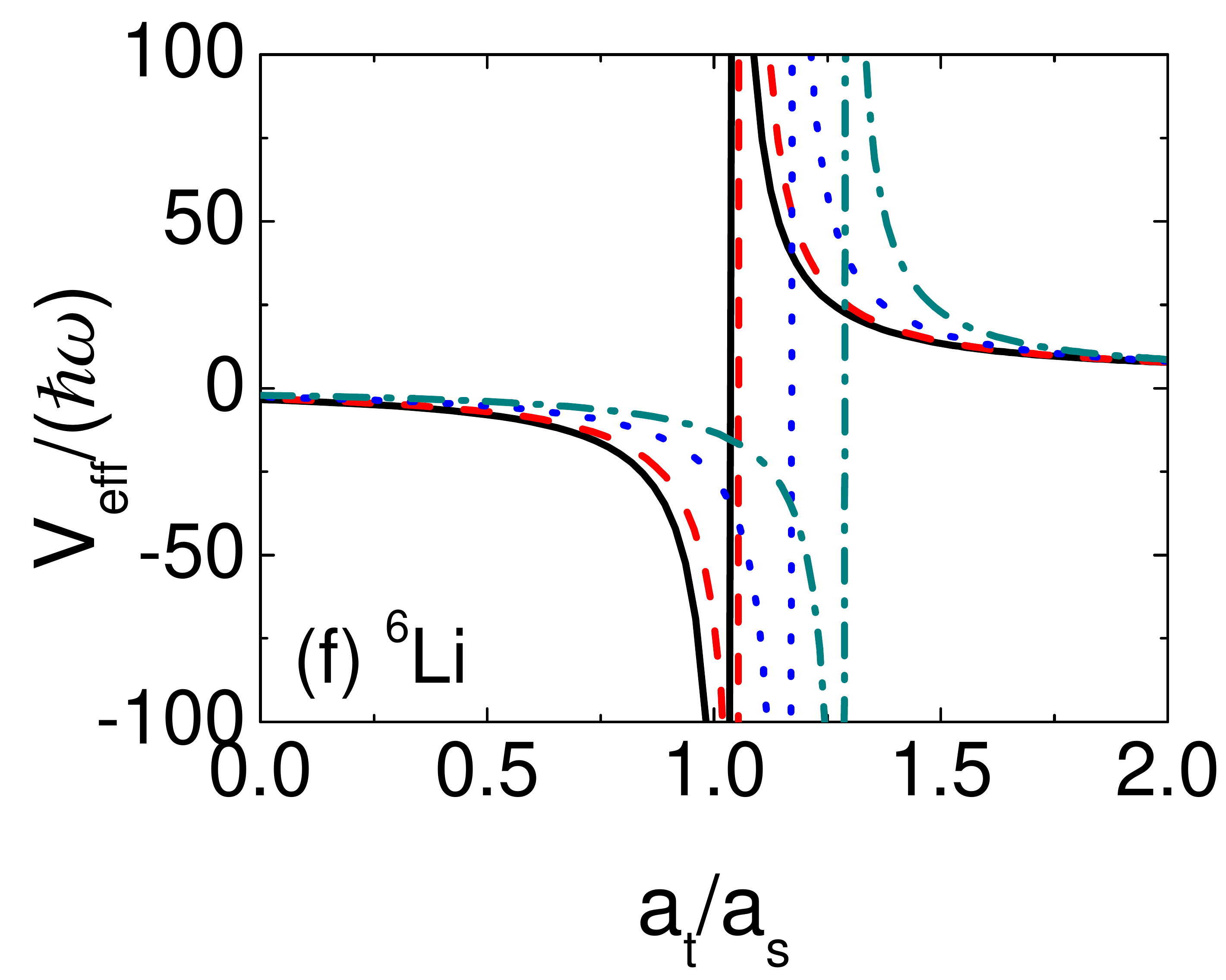}
\caption{Variation of parameters $\delta_p$, $\alpha_p$ and $V_{\rm eff}$ in 1D effective Hamiltonian as functions of $a_t/a_s$.}
\label{fig:effective-H}
\end{figure}

In Fig.~\ref{fig:effective-H}, we plot the parameters $\delta_p$, $\alpha_p$ and $V_{\rm eff} \equiv V_p-\alpha^2_p/(\delta_p-\varepsilon_b)$
for $^{40}{\rm K}$ and $^{6}{\rm Li}$ by tuning through a wide Feshbach resonance. The scattering parameters are taken as $W = 8$ G,
$a_{bg} = 174 a_B$, $\mu_{\rm co} = 1.68 \mu_B$ for $^{40}$K and $W = 300$ G, $a_{bg} = -1405 a_B$, $\mu_{\rm co} = 2 \mu_B$ for $^{6}$Li,
where $a_B$ is Bohr radius and $\mu_B$ represents Bohr magneton. With a typical transverse trapping frequency
$\omega_x = \omega_y = 2\pi \times 62$ kHz, the dimensionless physical parameters are then given by $g_p = 23$ $(272)$, $U_p = 1.7$ $(-5.5)$
for $^{40}$K ($^6$Li)~\cite{kestner,ren}. The qualitatively difference of $\delta_p$ and $\alpha_p$ is mainly due to the difference in sign of their
individual background interaction.

\section{Scattering states and the single-channel effective hamiltonian}
\label{sec:scattering states}

In this section, we focus on the energy regime above the threshold, and derive a 1D effective model by analyzing
the two-body scattering state. In this case, since a two-body bound state is absent, the transverse trapping potential
remains the largest energy scale such that the population of all transverse excited states are negligible as varying
the 3D scattering length. Thus, these high energy degrees of freedom can be dropped out when analyzing low-energy
physics, and a single-channel form is sufficient for a 1D effective Hamiltonian.

From Eq. (\ref{eqn:green1}), the behavior of $\langle0,0|g(\varepsilon_c;{\bf r,0})|0,0\rangle$ at vanishing $r$
associated with a two-body scattering state with energy $\varepsilon_c$ reads
\begin{eqnarray}
\label{eqn:green-vinish-r-scattering state}
\lim_{r\rightarrow0^+}\langle0,0|g(\varepsilon_c;{\bf r,0})|0,0\rangle
&=&\frac{1}{2\pi i \sqrt{\varepsilon_c}}-{\cal F}(\varepsilon_c),
\end{eqnarray}
where the function ${\cal F}$ is defined as in Eq. (\ref{eqn:f-function}). Following a similar approach as outlined in the previous section,
we obtain the contribution to the Green's function induced by the presence of SOC,
\begin{eqnarray}
\label{eqn:s-function-scatering state}
{\cal S}_s(\varepsilon_c)&=&\frac{1}{2\pi^2}\sum_{s=0}^{\infty}\int_{0}^{\infty} dx\frac{b\sqrt{x}}{bx+d}
\nonumber \\
&\times&\left[\frac{a_{s}+i0^+-x}{(a_{s}+i0^+-x)^2-(bx+d)}-\frac{1}{a_{s}+i0^+-x}\right].\nonumber
\\
\end{eqnarray}
By substituting Eqs.~(\ref{eqn:green-vinish-r-scattering state}), (\ref{eqn:s-function-scatering state}), and (\ref{eqn:scattering state}) into
the Bethe-Peierls boundary condition Eq.~(\ref{eqn:bp1}),  we can determine the coefficient $A(c)$
\begin{eqnarray}
\label{eqn:coefficient A}
A(c)&=&\frac{-2\pi^{3/2}\langle0,0|\psi_c^{(0)}(\textbf{0})\rangle}{- i \varepsilon_c^{-1/2} - \left({a_s}^{-1} - {\cal C}\right)
- \overline{\cal L}\left(-\varepsilon_c/2\right) / \sqrt{2} + 2\pi {\cal S}_s(\varepsilon_c)}
\end{eqnarray}
The scattering amplitude $f$ between the incident state $|\psi_c^{(0)}\rangle$ and the energy conserved output state $|\psi_{c'}^{(0)}\rangle$
is defined as
\begin{eqnarray}
\label{eqn:define scattering amplitude}
f(c'\leftarrow c)&=&-2\pi^2\langle\psi_{c'}^{(0)}(\textbf{0})|0,0\rangle A(c).
\end{eqnarray}
In the low-energy limit $\varepsilon_c \to \varepsilon_{\rm th}$, this 1D scattering amplitude can be expressed as
\begin{eqnarray}
\label{eqn:scattering amplitude}
f(c'\leftarrow c)&=&-\frac{1}{1+i\sqrt{\varepsilon_c}a_{\rm 1D}},
\end{eqnarray}
where the 1D scattering length takes the form
\begin{eqnarray}
\label{eqn:a1d}
a_{\rm 1D}&=&-\frac{a_t^2}{a_s}\left\{1-\frac{a_s}{a_t}\left[2\pi {\cal S}_s(\varepsilon_c)-\frac{\zeta\left(1/2,1-\varepsilon_c/2\right)}{\sqrt{2}}\right]\right\}.\nonumber\\
\end{eqnarray}
Here, we have brought back the unit of length to give a complete expression.

To derive a 1D effective Hamiltonian, we notice that the 1D scattering length, and hence the solution of two-body scattering states,
can be reproduced by considering a 1D pseudo-potential
\begin{eqnarray}
\label{eqn:U1d}
U_{\rm 1D}(x)&=&g_{\rm 1D}\delta(x),
\end{eqnarray}
where the coupling strength is
\begin{eqnarray}
\label{eqn:g1d}
g_{\rm 1D}(x)&=&-\frac{\hbar^2}{ma_{\rm 1D}}.
\end{eqnarray}
By further matching the single-particle dispersion, we obtain the 1D effective Hamiltonian
in a single-channel model form,
\begin{eqnarray}
\label{eqn:eff-hamilton}
\hat{H}_{\rm eff}^{\rm sc} &=&-\frac{\partial^2}{\partial x^2}+\lambda\sum_{j=1,2}\left[\frac{q_x}{2}+(-1)^{j}k_x\right]\sigma_{x}^{(j)}\nonumber\\
&+&h\sum_{j=1,2}\sigma_{z}^{(j)}+U_{\rm 1D}(x).
\end{eqnarray}
%


\section{confinement-induced resonance}
\label{sec:cir}

\begin{figure}
\includegraphics[width=7.0cm]{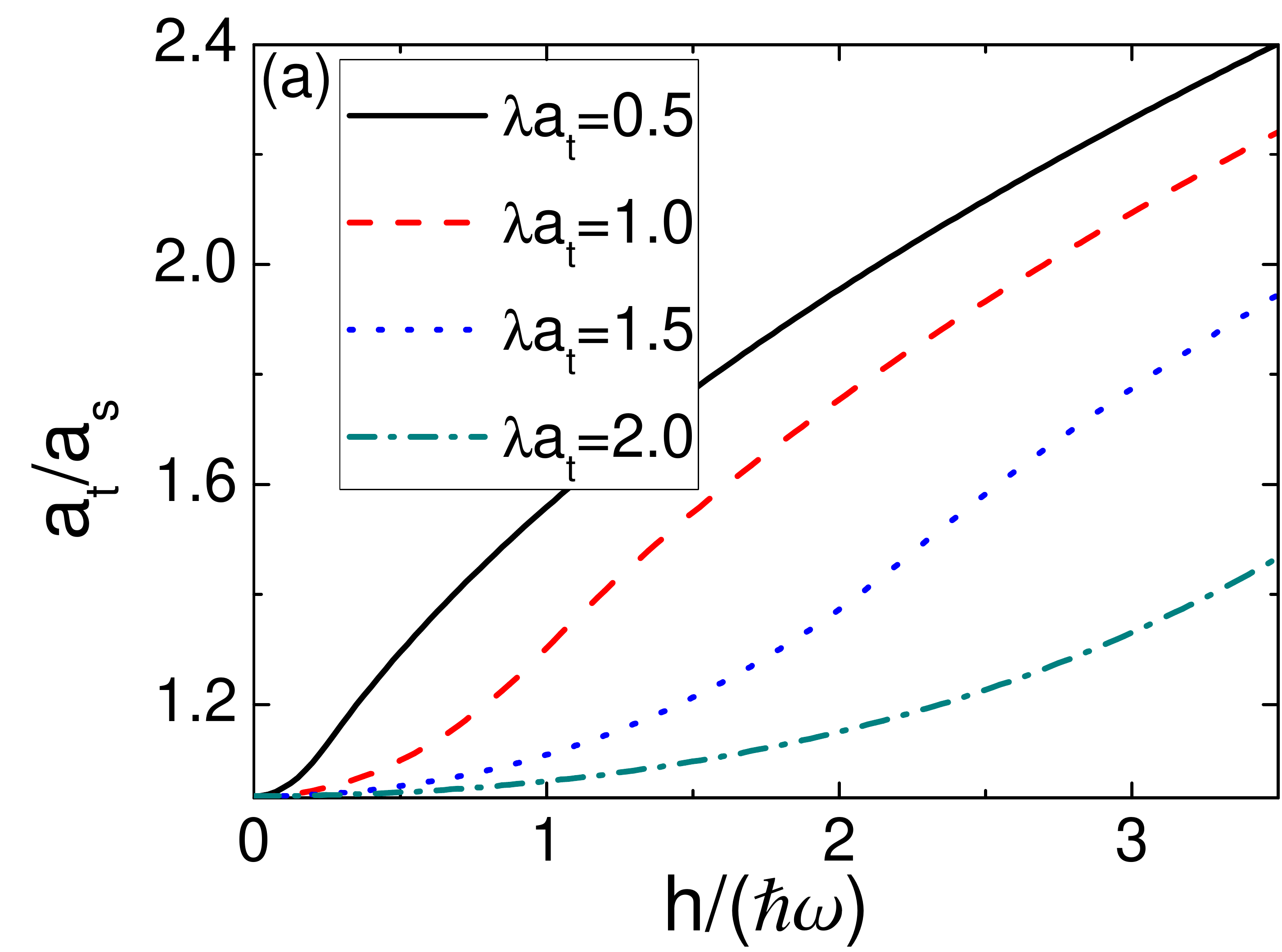}
\includegraphics[width=7.0cm]{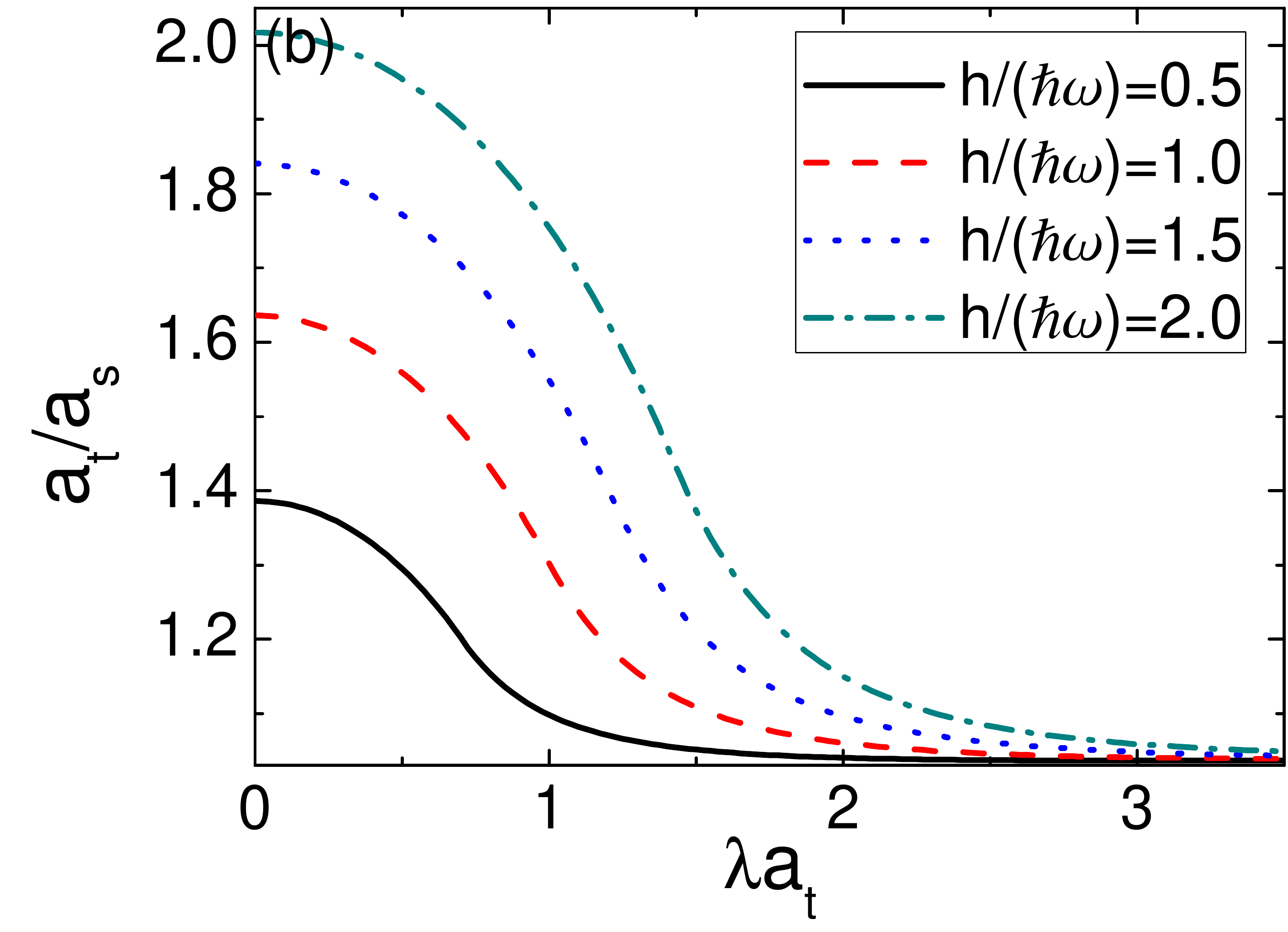}
\caption{Variation of CIR position $a_t/a_s$ as functions of (a) Zeeman field intensity $h$ and (b) SOC strength $\lambda$.}
\label{fig:cir-condition}
\end{figure}
The confinement-induced resonance is defined as a 1D resonance where the scattering amplitude $f = -1$, indicating a
complete reflection with $a_{\rm 1D} = 0$. Thus, the position for the CIR to take place can be derived from Eq. (\ref{eqn:a1d}),
leading to~\cite{yczhang-13}
\begin{eqnarray}
\label{eqn:cir}
\frac{a_t}{a_s}&=&2\pi{\cal S}_s(\varepsilon_{\rm th})-\frac{\zeta\left[1/2,1-\varepsilon_{\rm th}/2 \right]}{\sqrt{2}}.
\end{eqnarray}
In Fig.~\ref{fig:cir-condition}, we show the position of CIR by varying the intensities of  effective Zeeman field $h$ and SOC $\lambda$.
For a given SOC strength, $a_t/a_s$ increases monotonically with the effective Zeeman field, as shown in Fig.~\ref{fig:cir-condition}(a).
In the zero-field limit, since the SOC corresponds to an Abelian gauge field and can be dropped out from the problem via a unitary transformation,
the position of CIR reduces to the known result of $a_t/a_s = \zeta(1/2,1) \approx 1.0326$ for the case without SOC~\cite{olshanii-98}.
For a given effective Zeeman field, $a_t/a_s$ decreases monotonically with increasing SOC intensity, as shown in
Fig.~\ref{fig:cir-condition}(b). Notice that in the large SOC limit, the effect of Zeeman field becomes negligible, such that the position
of CIR takes the same value of $a_t/a_s = \zeta(1/2,1)$ as in the zero-field limit. On the other hand, as the SOC strength tends
zero, the position of CIR approaches to a limiting value, which is different from the result of the zero-SOC case,
and depends on the intensity of the effective Zeeman field. In fact, since SOC mixes the two spin states, the two-body threshold energy
will presents an abrupt change in the presence of an infinitesimally small SOC, leading to a finite shift of the position where
CIR takes place.

\begin{figure}
\includegraphics[width=7.0cm]{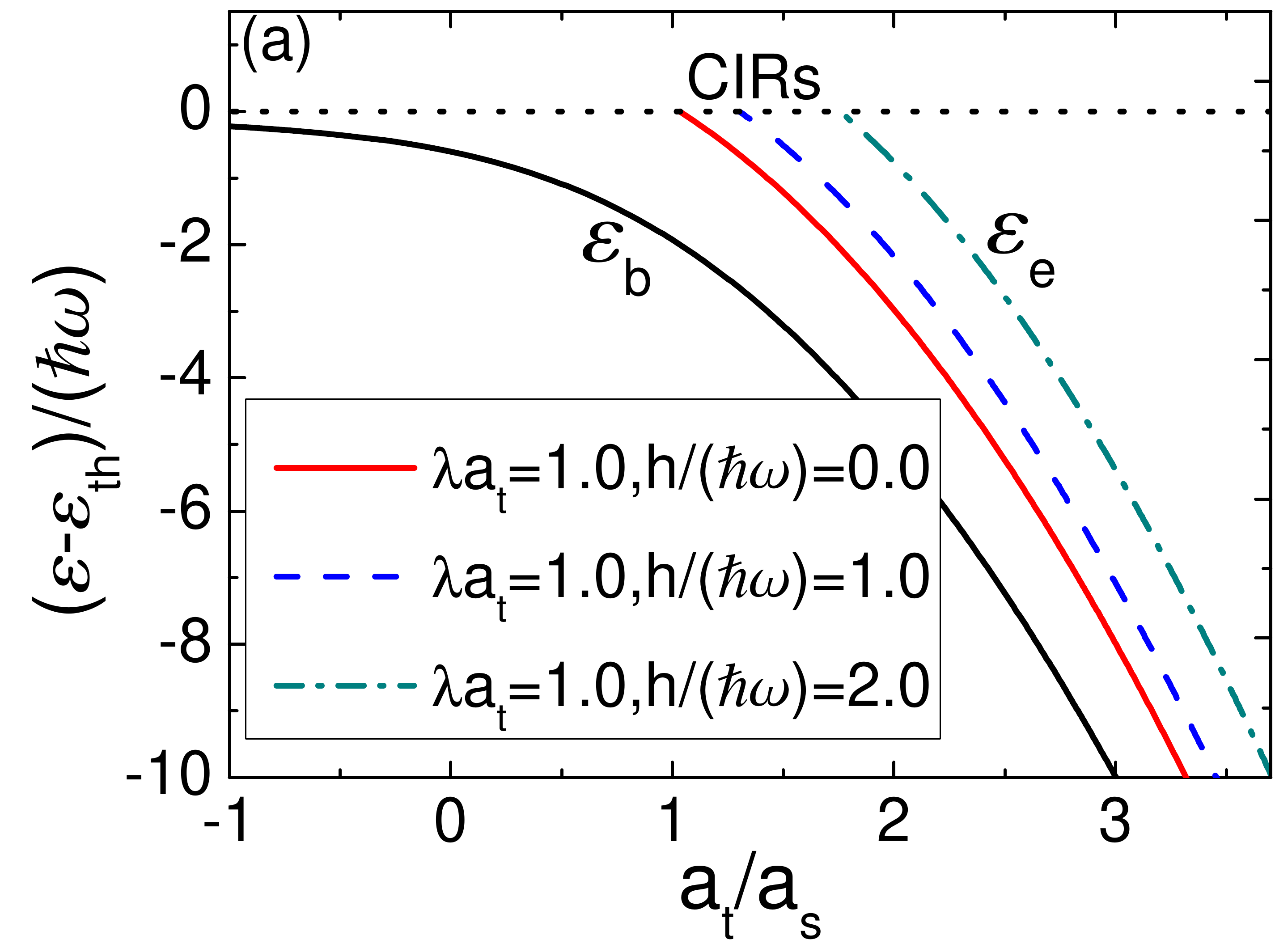}
\includegraphics[width=7.0cm]{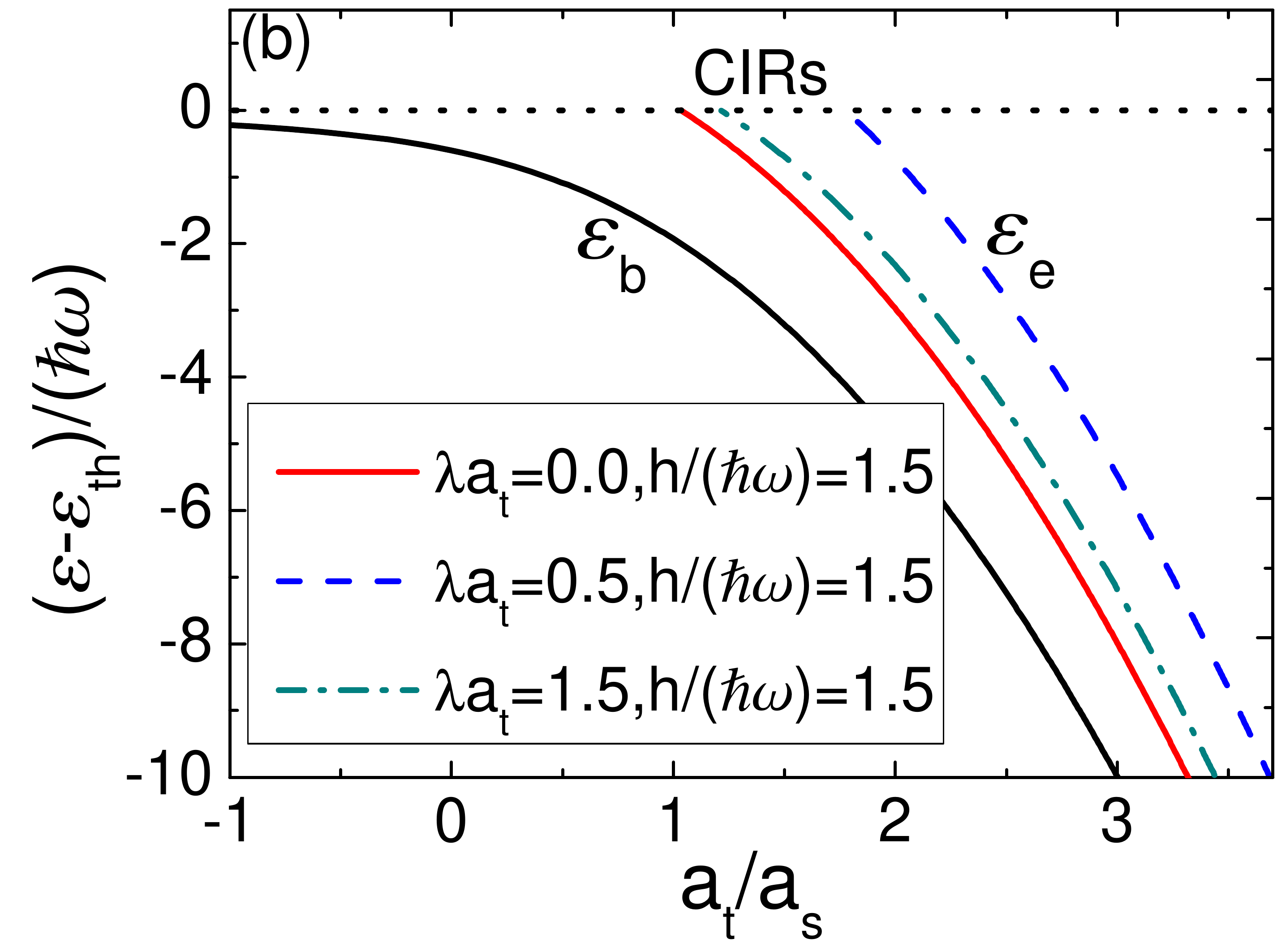}
\caption{Two-body bound state energies $\varepsilon_b$ and $\varepsilon_e$ as functions of $a_t/a_s$
by varying (a) the effective Zeeman field with a given SOC intensity and (b) the SOC intensity with a 
given effective Zeeman field. The crossing points of $\varepsilon_e$ and open channel threshold $\varepsilon_{\rm th}$ (dotted)
indicate the position of CIR to take place.}
\label{fig:bound-state}
\end{figure}

In the case without spin-orbit coupling, CIR can be understood as a Feshbach resonance which happens when the continuum threshold
of the open channel, which corresponds to the transverse ground state, degenerates with the two-body bound state energy of the closed channel,
which consists all transverse excited states~\cite{wzhang-11,olshanii-03}. In the presence of SOC, the two-body bound state energy
$\varepsilon_e$ within the closed channel can be derived using the same approach as in Sec.~\ref{sec:bound states}, leading to
\begin{eqnarray}
\label{eqn:Ee}
\frac{1}{a_s}&=&2\pi{\cal S}_e(\varepsilon_e)-\frac{\zeta\left[1/2,1-\varepsilon_e/2\right]}{\sqrt{2}},
\end{eqnarray}
where the function ${\cal S}_e$ takes the form
\begin{eqnarray}
\label{eqn:s-function-2}
&&{\cal S}_e(\varepsilon_e)
=
\frac{1}{2\pi}\sum_{s=1}^{\infty}\frac{b}{E_sb+d}
\nonumber \\
&& \times \left\{ \frac{\sqrt{2 \gamma-2 E_s-b}
\left[2 d-b\left(\gamma-E_s\right) \right]}{4 E_s b+b^2+4d}-\sqrt{-E_s}\right\}
\end{eqnarray}
with $\gamma = \sqrt{E_s^2 -d}$, $E_s = \varepsilon_e - 2s$, $b=4\lambda^2$ and $d=4h^2$.
In Fig.~\ref{fig:bound-state}, we plot the solution of $\varepsilon_e$ as a function of $a_t/a_s$ for
a various combination of SOC intensity and the strength of the effective Zeeman field.
From these results, we have confirmed that the crossing point of $\varepsilon_e$ and the open
channel threshold $\varepsilon_{\rm th}$ (denoted by dotted lines in Fig.~\ref{fig:bound-state}) exactly
coincides with the position of CIR. This observation indicates that the understanding of CIR as a
Feshbach resonance is still valid in the presence of SOC.
%


\section{summary}
\label{sec:con}
In this manuscript, we study the two-body problem of spin-1/2 fermionic atoms confined in a quasi-one-dimensional
harmonic trap with NIST-type spin-orbit coupling, and derive one-dimensional effective Hamiltonians for all energy regimes.
For energy regime close to the two-body bound state energy, since the transverse excited states can be
significantly populated at unitarity of even on the BCS side of a Feshbach resonance, the 1D effective model
takes a two-channel model form where the atoms in the transverse ground state assumes the open channel, while
the Feshbach molecules and the atoms in the transverse excited states are represented by structureless bosons
in the closed channel. The parameters in this effective Hamiltonian are fixed by matching single- and two-body physics,
This effective Hamiltonian can be used to investigate pairing physics within such a system where the fermionic chemical
potential is close to one half of the two-body bound state energy.

For energy slightly above the continuum threshold, the population of transverse excited states remains negligible such
that these degrees of freedom can be safely ignored when discussing low-energy behavior of the system.
Thus, we write down the 1D effective model in a single-model form, where the coupling constant of the 1D pseudo-potential
is determined by the 3D scattering length and transverse trapping potential. This model is useful when analyzing
the scattering processes or low-energy physics on the upper repulsive branch of a Feshbach resonance.
We also discuss the effect of SOC on the
position of confinement-induced resonances, where the system undergoes a complete reflection. Furthermore, we
show that in the presence of NIST-type SOC, CIR can still be understood as a Feshbach resonance where the
continuum threshold of the transverse ground state degenerates with the bound state energy of the closed channel
consisting of all transverse excited states. Considering the experimental realization of quasi-one-dimensionality
and SOC in fermionic systems, the shift of CIR position induced by SOC can be detected using
present technology at attainable temperatures.

\acknowledgments
We are grateful to Peng Zhang and Wei Yi for helpful discussion.
This work is supported by NKBRP (2013CB922000), NSFC (11274009),
and the Research Funds of Renmin University of China (10XNL016, 13XNH123).

\end{document}